\documentclass[a4paper]{jpconf}
\usepackage{graphicx}
\usepackage{xcolor}

\usepackage[T1]{fontenc} 
\usepackage{epsfig}  
\usepackage{graphicx}  
\usepackage{amssymb}
\usepackage{amsmath}
\usepackage{pifont}

\usepackage[english]{babel}
\usepackage{color}  
\definecolor{darkgreen}{rgb}{0.2,0.5, 0.2}

\usepackage{dcolumn,booktabs}
\newcolumntype{d}[1]{D{.}{.}{#1}}

\definecolor{wrcolor}{rgb}{0.0, 0.0, 0.0}
\newcommand*{\w}[1]{\textcolor{wrcolor}{#1}}

\definecolor{mbscolor}{rgb}{0.00, 0.0, 0.00}
\newcommand{\mb}[1]{\textcolor{mbscolor}{#1}}

\begin{document}

\title{The mass of odd-odd nuclei in microscopic mass models}

\author{W. Ryssens$^1$, G. Scamps$^2$, G. Grams$^1$, I. Kullmann$^1$, M. Bender$^{3}$ and S. Goriely$^1$.}
\address{$^{1}$Institut d’Astronomie et d’Astrophysique, Université Libre de Bruxelles, Campus de la Plaine CP 226, 1050 Brussels, Belgium}
\address{$^{2}$Department of Physics, University of Washington, Seattle, Washington 98195-1560, USA}
\address{$^{3}$Universit{\'e} de Lyon, Universit{\'e} Claude Bernard Lyon 1, CNRS, IP2I Lyon / IN2P3, UMR 5822, F-69622, Villeurbanne, France \label{addr3}
}

\ead{wouter.ryssens@ulb.ac.be}

\begin{abstract}
Accurate estimates of the binding energy of nuclei far from stability 
that cannot be produced in the laboratory are crucial to our understanding of nuclear
processes in astrophysical scenarios. Models based on energy density 
functionals have shown that they are capable of reproducing all known masses
with root-mean-square error better than 800 keV, while retaining a firm
microscopic foundation. 
However, it was recently pointed out in 
[M. Hukkanen et al., arXiv:2210.10674] that the recent BSkG1 model fails to 
account for a contribution to the binding energy that is specific
to odd-odd nuclei, and which can be studied \w{by using} appropriate mass 
difference formulas. We analyse here the (lacking) performance of three 
recent microscopic mass models with respect to such formulas and examine
possibilities to remedy this deficiency in the future.
\end{abstract}

\section{Introduction}

The modelling of nuclear processes that impact stellar evolution and nucleosynthesis
relies on our knowledge of the properties of atomic nuclei. Astrophysical applications
require reliable values for many different quantities, but the central observable
in this context is the nuclear mass or binding energy~\cite{Arnould20}. 
For many relevant nuclei far from stability, however, 
the experimental difficulties of producing and handling unstable 
isotopes preclude mass measurements in the foreseeable future. 

This knowledge gap can only be filled through the development of models,
which should provide reliable extrapolations for the masses of exotic nuclei. Our
confidence in such extrapolations naturally increases if such models can 
provide a faithful description of known experimental data, not only for nuclear 
masses but also with respect to other quantities such as charge radii and 
nuclear deformations. Models based on energy density functionals (EDFs) 
\cite{Bender03} are the most promising avenue in this respect, since they 
have a microscopic foundation in an effective (pseudo-)potential between individual 
nucleons while their application to thousands of nuclei remains feasible. However, 
the analytical form of the EDF is phenomenological and 
its coupling constants are adjusted to experimental data. 
Many forms of the EDF and parameter adjustment strategies have been 
proposed~\cite{Bender03}. We focus here on those developed by the Brussels group:
the BSk~\cite{Goriely16} and BSkG~\cite{Scamps21,Ryssens22} 
series are based on EDFs of the Skyrme type and are fitted to essentially all 
known nuclear masses. These models describe the known nuclear masses with 
a root-mean-square (rms) error that is below 0.8 MeV~\cite{Goriely16,Scamps21,Ryssens22}, 
an accuracy that is competitive with more phenomenological microscopic-macroscopic approaches~\cite{Moller16}.

Experimentally, the mass of a nucleus with an odd number of neutrons $N$ and \w{an}
odd number of protons $Z$ is somewhat smaller than would naively be expected 
from the masses of its neighbouring even-even, odd-$Z$ and odd-$N$ isotopes. 
This additional binding energy is visible in the 
systematics of mass differences and is often taken as evidence for 
the existence of a residual interaction between the odd proton and odd neutron.
This effect has been discussed at length in the literature, see for example \cite{Basu71,Jensen84,Friedman07,Robledo14,Wu16,Yang22}, 
but without reaching a consensus about its microscopic origin and nature. We
ask here the pragmatic
question: do microscopic mass models account for the effect of this interaction
on the binding energies of odd-odd nuclei? Through the use of appropriate
mass difference formulas, we will show that the answer is "no", confirming 
the discussions in Ref.~\cite{Hukkanen22,Ryssens22}. We discuss possibilities 
to improve future models by accounting for this effect in a microscopic way.

\section{Mass differences}
\label{sec:residual}

Leaving aside shell effects, the (positive)
binding energies $B(N,Z)$ of even-even atomic 
nuclei are rather smooth functions of both neutron number $N$ and proton number 
$Z$. Adjacent odd-mass nuclei follow almost identical trends with nucleon number, but are systematically less bound than the average of their 
even-even neighbours, an effect that is attributed to the presence of pair correlations in nuclei \cite{Bender00,Duguet01b}. Using $\Delta_q$ ($q=p$, $n$) 
for the distance between the curves that interpolate the masses of even-even and odd-mass nuclei, this can be schematically written as
\begin{align}
B(N,Z) & 
\simeq \tfrac{1}{2} \left[ B(N-1,Z) +  B(N+1,Z)\right] + (-1)^{N} \Delta_n \, ,
\label{eq:oddN}
\\
B(N,Z) & 
\simeq \tfrac{1}{2} \left[ B(N,Z-1) +  B(N,Z+1)\right] + (-1)^{Z} \Delta_p \, ,
\label{eq:oddZ}
\end{align}
where $\Delta_n$ and $\Delta_p$ vary only slowly with particle number when away 
from closed shells. Values of these shifts can be extracted approximately from 
experiment with suitable mass difference formulas, such as the three- and five-point gaps $\Delta^{(3)}_{q}$
and $\Delta^{(5)}_{q}$~\cite{Bender00,Duguet01b}. The experimental values determined 
this way typically range between 0.5 and 1.5 MeV.


The evolution of the binding energy of odd-odd nuclei with nucleon number is 
also similar to that of the even-even nuclei. However, the distance between 
the mass surface of even-even and odd-odd nuclei is generally \emph{smaller} 
than $\Delta_n + \Delta_p$, which is what one would expect naively from 
Eqs.~\eqref{eq:oddN} and \eqref{eq:oddZ}. Schematically, we write for odd values
of $N$ and $Z$:
\begin{align}
\label{eq:def_dnp} 
B(N,Z) 
       &\simeq  \tfrac{1}{2} \left[ B(N+1,Z-1) + B(N-1,Z+1)\right] - \Delta_n  - \Delta_p
                 + \Delta_{np} \, . 
\end{align}
As we will show below, measured nuclear masses systematically point to values of
$\Delta_{np}$ that are on the order of a few hundred keV. 
%
%
Different mass formulas have been suggested to \w{quantify this effect based on the} 
measured masses~\cite{Wu16,Yang22}, but we focus on the following {quantity:
\begin{align}
\Delta^{(3)}_{np}(N,Z) &= (-1)^{Z}   \bigg[\Delta^{(3)}_n(N,Z) - \Delta^{(3)}_n(N,Z-1) \bigg] =   -(-1)^{Z+N} S_{p2n} \, .
\label{eq:d3np}
\end{align}
where $S_{p2n}$ characterizes the odd-even staggering of the proton separation
energy $S_p$ along an isotonic chain, i.e.\ 
$S_{p2n}(N,Z) \equiv S_p(N,Z) - \tfrac{1}{2} [ S_p(N+1,Z) + S_p(N-1,Z)]$~\cite{Friedman07}.
An alternative definition of $\Delta^{(3)}_{np}$ in terms of proton three-point 
gaps energies leads to similar conclusions. 
$\Delta^{(3)}_{np}$ is closely related, but not identical, to the quantity 
$\delta_{np}$ studied in Ref.~\cite{Wu16}. 

In the left panel of Fig.~\ref{fig:dpn_exp}, we show the $\Delta^{(3)}_{np}$ values 
calculated from the experimental masses tabulated in the AME20 compilation~\cite{Wan21}.
Essentially all values are positive: of the 1993 \w{$(N,Z)$ pairs}
for which we can evaluate $\Delta^{(3)}_{np}$, only 28 result in negative values. 
Positive values of $\Delta^{(3)}_{np}$ indicate that the three-point neutron gaps along 
odd-$Z$ isotopic chains are smaller than the gaps along neighbouring even-$Z$ chains.
We observe several outliers larger than
one MeV, but these occur almost exclusively for nuclei with
$|N - Z| \leq 1$, whose binding energies follow more complex patterns 
than those outlined above~\cite{Yang22}. For all other nuclei,  the typical
effect amounts to a few hundred keV, decreasing slowly with increasing \w{mass}.
The overall trend can be fit\w{ted} with a simple $A$-dependence~\cite{Basu71,Friedman07,Yang22},
but the large scatter seems to indicate that shell structure also plays an
important role. For the purpose of simple comparison, the average of all 
experimental values is 305 keV.

As pointed out in Ref.~\cite{Hukkanen22} for the example of Ru, Rh and Pd isotopes,
this empirical trend of the $\Delta^{(3)}_{np}$ is not necessarily described by microscopic 
mass models. To illustrate how this deficiency is already visible for a more 
conventional mass difference, the right panel of Fig.~\ref{fig:dpn_exp} shows the 
proton separation energies along 
the Nd, Gd, Yb, W and Hg isotopic chains. BSkG1 produces rather smooth curves that
reproduce the overall trends of experimental data, but the clearly
visible odd-even staggering of the experimental data is absent from the model.
Similar conclusions apply \w{to} neutron separation energies along isotonic chains.
The actual $\Delta^{(3)}_{np}$ values obtained from the BSkG1 model
(for the same set of nuclei as in Fig.~\ref{fig:dpn_exp}) are shown in the 
top left panel of Fig.~\ref{fig:dpn}. The difference with experimental 
information is obvious: about half of all calculated values are negative 
and the average value is very close to zero. 

\begin{figure}
\centering
\includegraphics[width=.425\textwidth]{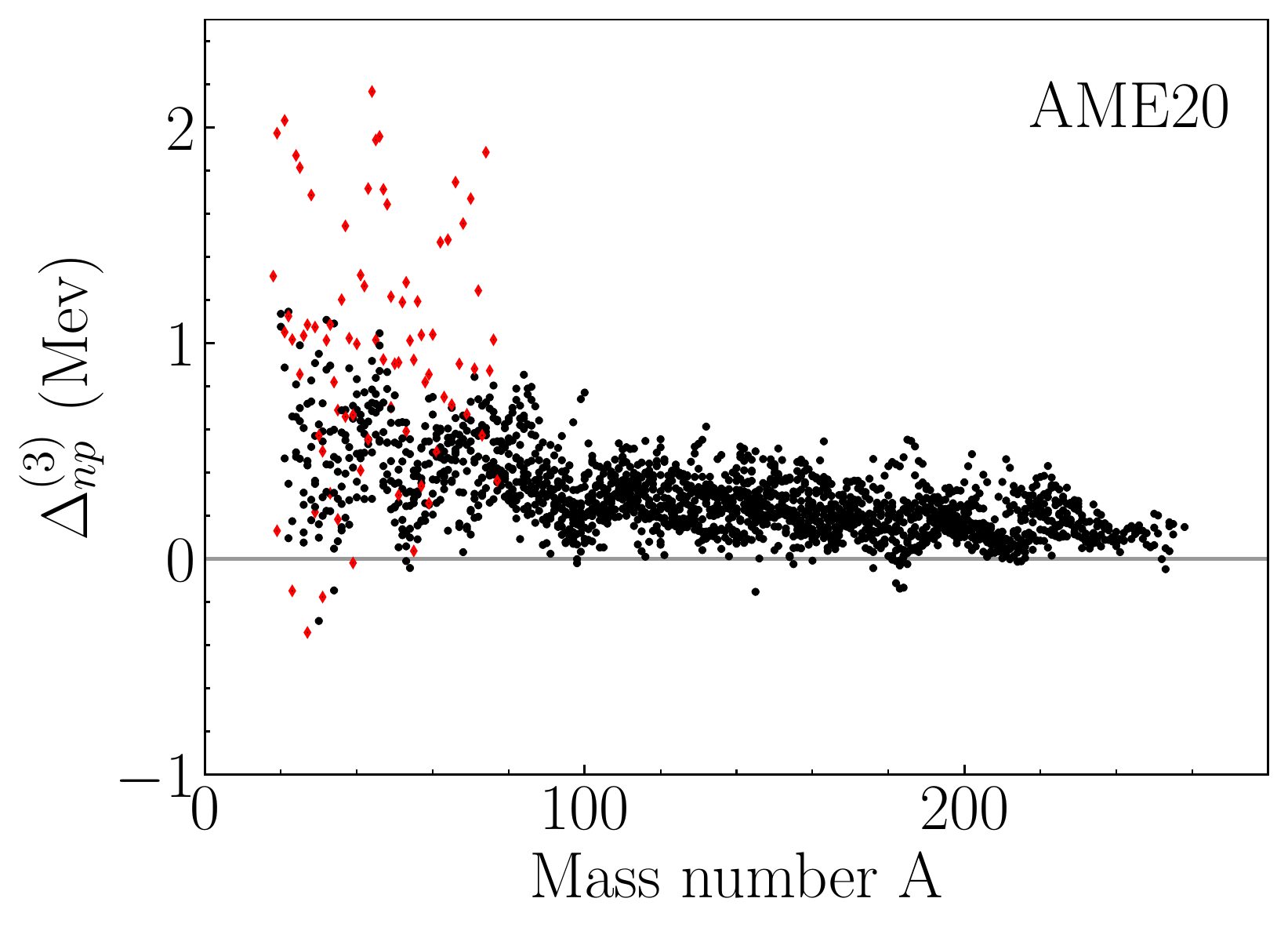}
\includegraphics[width=.425\textwidth]{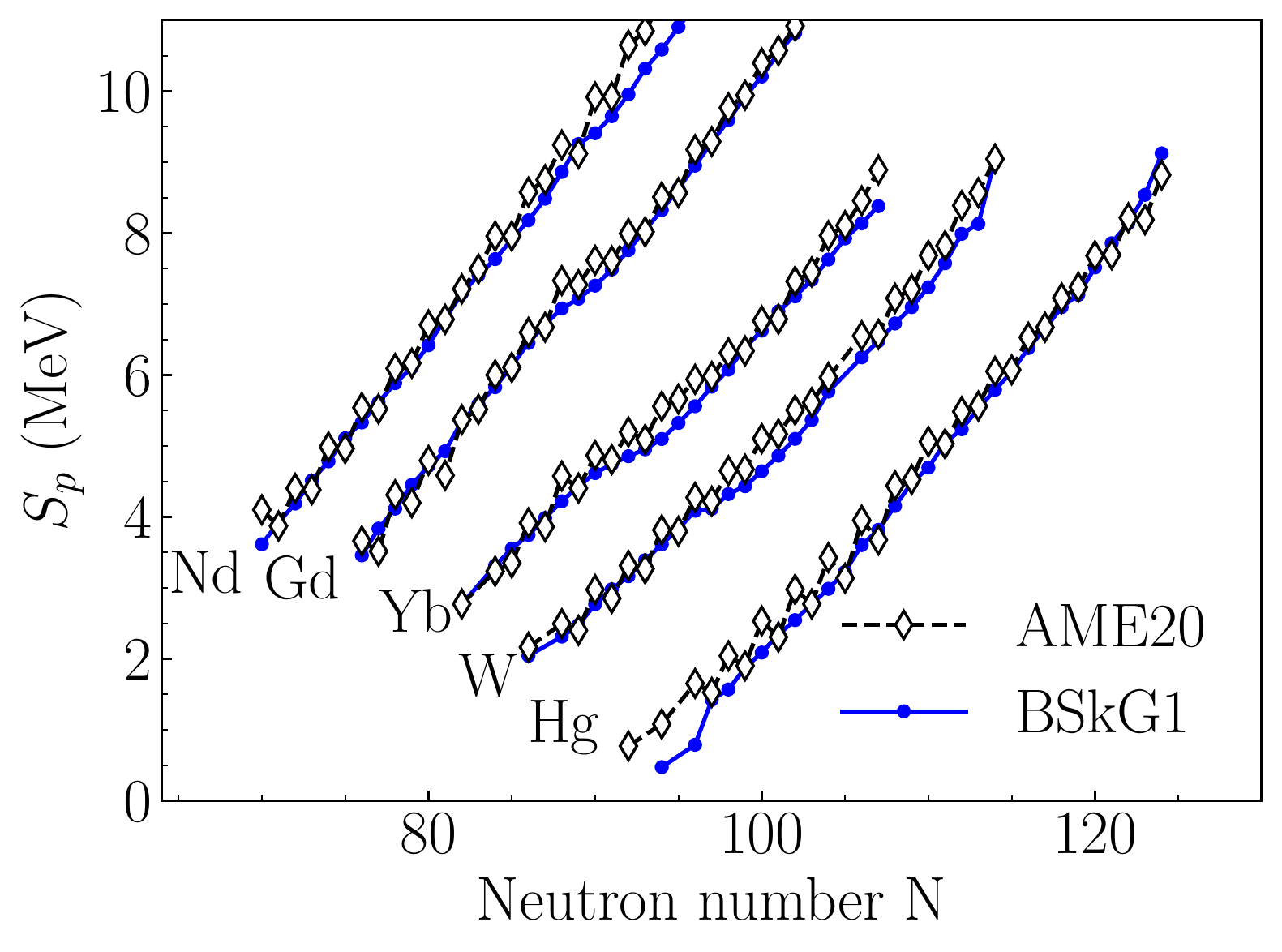}
\caption{(Color online) Left: $\Delta^{(3)}_{np}$ values for nuclei with $Z\geq 8$
         calculated from the known masses~\cite{Wan21}. Red diamonds indicate nuclei with $|N-Z| \leq 1$.
         Right: Proton separation energies $S_p$ along the Nd, Gd, Yb, W and Hg
         isotopic chains. We show experimental values (empty black diamonds)~\cite{Wan21} and 
         BSkG1 values (blue circles). 
}
\label{fig:dpn_exp}
\end{figure}

\begin{figure}
\centering
\includegraphics[width=.75\textwidth]{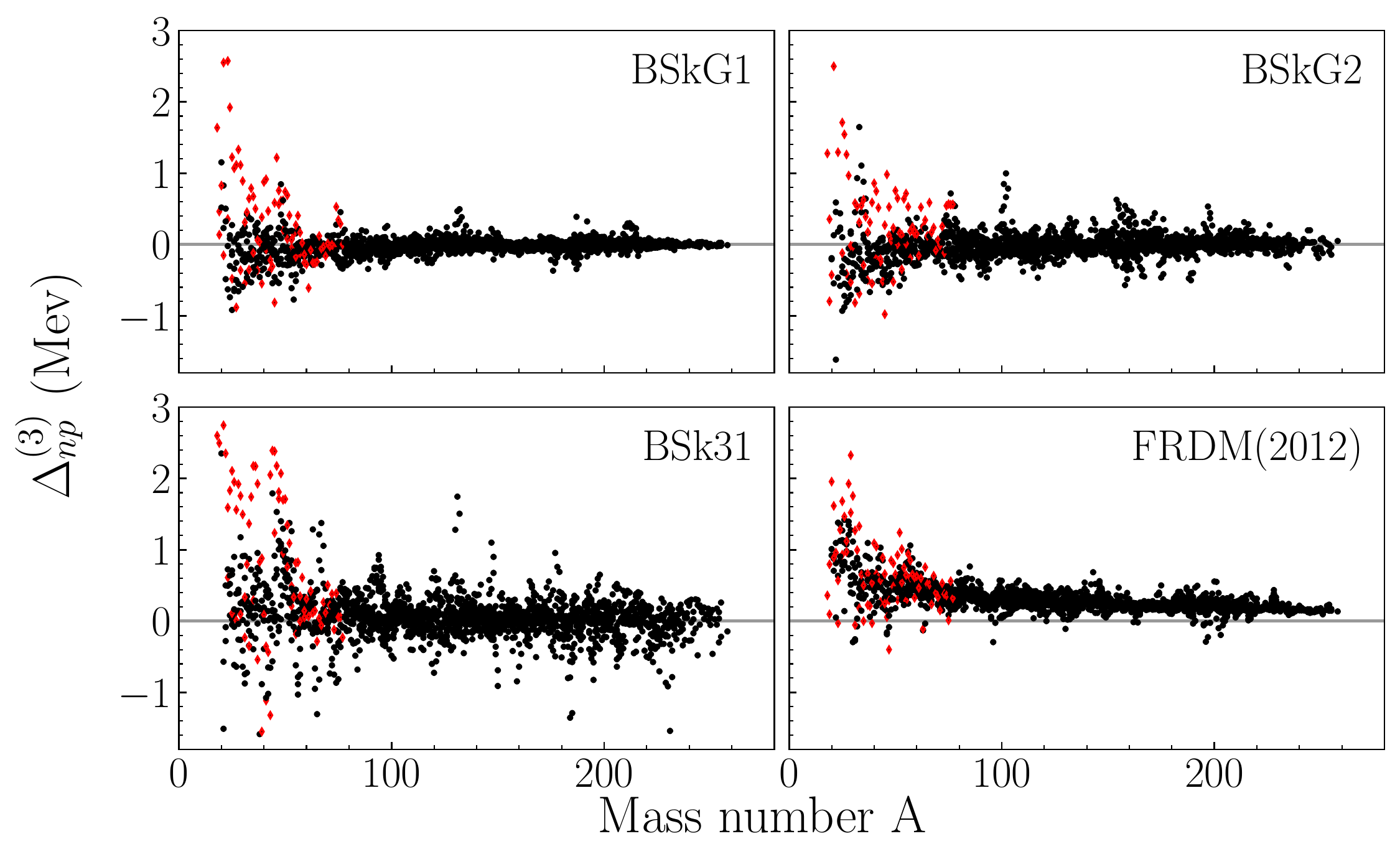}
\caption{(Color online) $\Delta^{(3)}_{\w{np}}$ values for nuclei with $Z\geq 8$
         whose masses are known~\cite{Wan21}
         calculated from four different mass models: BSkG1 (top left), BSkG2
         (top right), BSk31 (bottom left) and FRDM(2012) (bottom right).
         Red diamonds indicate nuclei with $|N-Z| \leq 1$.
}
\label{fig:dpn}
\end{figure}

Other microscopic mass models face similar problems, which is illustrated 
in Fig.~\ref{fig:dpn} for BSkG2~\cite{Ryssens22} and BSk31~\cite{Goriely16}.
We also compare with the macroscopic-microscopic 
FRDM(2012) model~\cite{Moller16} that} explicitly includes a phenomenological 
term proportional to $A^{-2/3}$ for odd-odd nuclei that models $\Delta_{np}$.
All models we consider offer an excellent global description of AME20 masses~\cite{Wan21}: 
the rms deviations for BSkG1 and BSkG2 are 741 and 678 keV respectively, 
while BSk31 and FRDM(2012) achieve slightly smaller values, 588 and 606 keV,
respectively. 

To the best of our knowledge, the recent BSkG2 is the only microscopic mass model 
so far that consistently included the so-called `time-odd' terms during the parameter 
adjustment. \w{In principle,} said terms contribute to the binding energy of odd-mass and odd-odd nuclei, but
are omitted by all other models through the means of the equal filling 
approximation~\cite{Perez08}, which greatly simplifies the solution of the 
mean-field equations. Even though this approximation is, in general, quite
accurate~\cite{Ryssens22}, one might expect that BSkG2 achieves an 
above-standard description of the binding energies of odd-odd nuclei.

\w{We also show results for BSk31, which differs from the more recent BSkG-series in several ways. 
Most of these differences do not directly impact our discussion, except for the approach to
the proton and neutron pairing strengths.}
\w{Both BSkG1 and BSkG2} employ a single parameter to characterize the pairing 
strength for each nucleon species. For reasons outlined in Ref.~\cite{Nayak95},
BSk31 generalizes this approach by taking two parameters for each species, 
depending on whether the corresponding particle number is even or odd. 
Through this mechanism, the proton (neutron) pairing strength in odd-$Z$ (odd-$N$) 
nuclei is enhanced
by 6\% (4\%) compared to even-$Z$ (even-$N$) ones. Both pairing strengths are
increased in odd-odd systems, providing the possibility for additional 
binding energy in these nuclei~\cite{Nayak95}.

The $\Delta^{(3)}_{np}$ values obtained from these models are shown in 
Fig.~\ref{fig:dpn}. As expected, FRDM(2012) describes the overall size and 
trend of the experimental data; the 
average calculated $\Delta^{(3)}_{np}$ among this set of nuclei is about 
306 keV.
The microscopic models fare much worse: in all of them, about half of the
values come out negative, resulting in average values of $\Delta^{(3)}_{np}$ of 
almost precisely 0 keV for BSkG1 and BSkG2, respectively, and \w{about 100 keV
for BSk31}. \w{While the calculated average $\Delta^{(3)}_{np}$ is at least 
positive for BSk31, this model} yields stronger scattering than BSkG1 and BSkG2, 
which are closer to experimental \w{data} in this respect. 

For the BSkG2 and BSk31 models, the rms error on the 585 known masses of
 odd-odd nuclei amounts to 704 and 616 keV, i.e.\ about 40 keV above their 
 respective rms deviation on  all masses. It is tempting to interpret 
 this as a consequence of the missing physics discussed here. However, 
 model deficiencies that affect a subset of nuclei feed back on the parameter 
 adjustment which targets an optimal description of all masses. The incomplete
 description of the binding energy of odd-odd nuclei thus need not result
 in an increased rms for such systems. For example, the BSkG1 model has an rms 
 deviation for odd-odd nuclei that is equal to its rms error on all masses.

\section{Perspectives}

We have established that BSkG1, BSkG2 and BSk31 do not reproduce the enhanced
binding energies of odd-odd nuclei observed experimentally. For BSkG1, there is
no clear mechanism that could have produced the desired effect. BSkG2 and
BSk31 provide a similar (lacking) description of the data, despite the presence of 
time-odd terms in the former and the alternating pairing strength in the latter.
This conclusion comes with the caveat that none of these models included
observables in their fitting protocols that are sensitive to this
small effect.

Although we only show results for three models, we are under the 
impression \w{that the failure to describe this additional binding of odd-odd nuclei is universal among EDF-based approaches}. 
There are several possible reasons \w{for this}. 
\w{First,}
\w{the form of the EDF might} not contain the necessary degree of freedom needed 
to describe \w{the additional binding energy}. \w{Even if the structure of the EDF would be sufficiently rich, 
the} fit protocol might not constrain this specific \w{mass} difference
or \w{the effect} might be related to a \w{broken symmetry that is not explored in the calculations}. \mb{Finally, there}
here also is the possibility that the modeling \mb{of} this effect requires \w{considering} a specific type 
of correlations in a beyond-mean-field approach.

\mb{The failure we observe might be due to a combination 
of all of the above,} 
but we will limit our considerations to modifications 
of the current state-of-the\w{-}art models that could generate additional binding 
energy in odd-odd nuclei without abandoning the mean-field description.
\w{It is likely that} the problem \w{we discuss here} is linked to another 
known problem: \w{EDF-based models do not reproduce} \w{the} energy splitting of states in odd-odd 
nuclei that are obtained from coupling the same two quasiparticles with either aligned or anti-aligned spins~\cite{Robledo14}. 
Whatever is the mechanism
that generates the extra binding \w{energy}, it is clear that it has to be sensitive to
the relative orientation of the angular momenta of the two odd nucleons and has 
to favour the parallel alignment of \w{their} intrinsic spin\w{s} as suggested
by the Gallagher-Moszkowski (GM) rule~\cite{Robledo14}.


One mechanism at the origin of the extra binding might be the formation of 
spin-aligned proton-neutron pairs in odd-odd nuclei, whose mean-field description requires 
breaking the conserved isospin projection of the single-particle states and 
the introduction of proton-neutron pairing \w{terms in the} EDF. As argued in 
Ref.~\cite{Friedman07} however, this is unlikely to generate the desired effect 
in a pure mean-field approach, where the formation of a static neutron-proton pair 
condensate usually only occurs near the $N=Z$ line. 
On the other hand, it is also known
that dynamical proton-neutron pairing is a necessary ingredient for the QRPA modeling 
of the $\beta$-decay of even-even to odd-odd nuclei across the chart of nuclei \cite{Engel99}, \mb{pointing to an inherent limitation of the mean-field approach to capture all relevant pairing correlations simultaneously.}
%

If the modeling of the enhanced binding of odd-odd nuclei is a beyond-mean-field 
effect, \w{one might attempt to include it through a phenomenological correction} 
similar to the ones \w{for collective motion} that are employed in 
the BSk and BSkG models \cite{Scamps21,Ryssens22,Goriely16}.
The simplest possibility is the adoption of an 
analytical formula such as the one included in the FRDM(2012) model, 
which works well for the global trends of the known nuclear masses. 
More microscopic correction formulas in the form of sums over matrix elements of
a residual interaction can be designed as well~\cite{Friedman07}, \w{but might 
double-count contributions} already \w{included} in the EDF.

This brings us to a third possibility: tuning the 
proton-neutron spin-spin interaction terms in the Skyrme EDF that only contribute 
when time-reversal symmetry is broken, meaning that it vanishes for even-even nuclei. 
The most simple one among these terms contains the scalar product of the proton 
and neutron spin densities 
$\mathbf{s}_p(\mathbf{r}) \cdot \mathbf{s}_n(\mathbf{r})$~\cite{Ryssens22}. 
The spin densities encountered in ground-state configurations are usually 
dominated by the single-particle contribution of the odd nucleons,\footnote{Polarisation 
effects will also induce a small spin density for the nucleons of the other
species, such that the spin density of the nucleon species with even number
will not vanish exactly in odd-mass nuclei.} 
such that terms involving $\mathbf{s}_n(\mathbf{r}) \cdot \mathbf{s}_p(\mathbf{r})$ will 
only significantly impact the ground states of odd-odd nuclei. The scalar 
product of the pseudovector spin densities also introduces a dependence 
on the relative orientation of the nucleon's spins as expected from the GM rule.
The fine-tuning of the spin-spin terms to the enhanced binding of odd-odd nuclei,
however, should not be made at the expense of
other qualities of the model. For example, changes in the proton-neutron 
spin terms might spoil the realistic values of the Landau parameters $G_0$ and $G_0'$ 
that BSkG1, BSkG2 andd BSk31 all exhibit, which then would call for higher-order
spin terms such as tensor interactions that were already evoked in earlier attempts 
to model the GM splitting~\cite{boisson76}.
In any case, since BSkG2 already incorporates time-odd terms \w{yet fails to describe 
the experimental values of $\Delta^{(3)}_{np}$}, \w{progress along this line will 
require amending the fit protocol with suitable observables}.



\section{Conclusions}

The binding energy of odd-odd nuclei is slightly, but systematically, larger
than what would be expected from estimates based on the masses of 
their even-even and odd-mass neighbours.
We have investigated
whether large-scale microscopic mass models describe this effect through the 
$\Delta^{(3)}_{np}$ mass difference.
As could be suspected from the discussion in Ref.~\cite{Hukkanen22}, 
the BSkG1 model produces values for this observable that are scattered 
around zero while experiment indicates systematically positive values. 
BSk31 and BSkG2 possess mechanisms that could generate additional binding 
energy in odd-odd nuclei, but in practice do not offer a better description of
the data. The macroscopic-microscopic FRDM(2012) 
model employs a phenomenological ingredient that explicitly accounts for the additional binding.
We have discussed perspectives to improve future microscopic models and, 
for models based on EDFs of the Skyrme type, have pointed out that specific 
time-odd terms could be used to generate additional binding energy in 
odd-odd nuclei.

\section*{Acknowledgments}
This work was supported by the Fonds de la Recherche Scientifique (F.R.S.-FNRS) 
and the Fonds Wetenschappelijk Onderzoek-Vlaanderen (FWO) under the EOS 
Project nr O022818F.
Computational resources have been provided by the Consortium des Équipements de Calcul Intensif (CÉCI), funded by the Fonds de la Recherche Scientifique de Belgique (F.R.S.-FNRS) under Grant No. 2.5020.11 and by the Walloon Region

The present research benefited from computational resources 
made available on the Tier-1 supercomputer of the F\'ed\'eration 
Wallonie-Bruxelles, infrastructure funded by the Walloon Region under the grant 
agreement nr 1117545. 
The funding for G.S. from the US DOE, Office of Science, 
Grant No. DE-FG02- 97ER41014 is greatly appreciated. S.G.\ and W.R.\ acknowledge
financial support from the F.R.S.-FNRS (Belgium). Work by M.B.\ has been 
supported by the French Agence Nationale de la Recherche under grant 
No.\ 19-CE31-0015-01 (NEWFUN). 

\end{document}